\documentclass[11pt]{article}
\usepackage{amsmath,amssymb,graphics,cite}
\catcode`\@=11

\@addtoreset{equation}{section}
\def\theequation{\arabic{section}.\arabic{equation}}
     
\catcode`\@=11
\def\thesection{\arabic{section}.}

\def\appendix{\setcounter{section}{0}
        \def\thesection{Appendix.}
        \def\theequation{\Alph{section}.\arabic{equation}}}
\def\section{\@startsection{section}{1}{\z@}{3.5ex plus 1ex minus
   .2ex}{2.3ex plus .2ex}{\large\bf}}

\long\def\@makefntext#1{\parindent 0cm\noindent
\hbox to 1em{\hss$^{\@thefnmark}$}#1}

\newcommand{\captionfonts}{\small}
\makeatletter  
\long\def\@makecaption#1#2{%
  \vskip\abovecaptionskip
  \sbox\@tempboxa{{\captionfonts #1: #2}}%
  \ifdim \wd\@tempboxa >\hsize
    {\captionfonts #1: #2\par}
  \else
    \hbox to\hsize{\hfil\box\@tempboxa\hfil}%
  \fi
  \vskip\belowcaptionskip}
\makeatother   

\begin{document}
\begin{titlepage}
\vspace{.5in}
\begin{flushright}
UCD-05-11\\
gr-qc/0508071\\
August 2005\\
\end{flushright}
\vspace{.5in}
\begin{center}
{\Large\bf
 Horizon constraints and black hole entropy}\\
\vspace{.4in}
{S.~C{\sc arlip}\footnote{\it email: carlip@physics.ucdavis.edu}\\
       {\small\it Department of Physics}\\
       {\small\it University of California}\\
       {\small\it Davis, CA 95616}\\{\small\it USA}}
\end{center}

\vspace{.5in}
\begin{center}
{\large\bf Abstract}
\end{center}
\begin{center}
\begin{minipage}{4.75in}
{\small
To ask a question about a black hole in quantum gravity, one must 
restrict initial or boundary data to ensure that a black hole is
actually present.  For two-dimensional dilaton gravity, and probably
a much wider class of theories as well, the imposition of a ``stretched 
horizon'' constraint alters the algebra of symmetries at the horizon, 
introducing a central term.  Standard conformal field theory techniques 
can then then be used to obtain the asymptotic density of states, 
reproducing the Bekenstein-Hawking entropy.  The microscopic states 
responsible for black hole entropy can thus be viewed as ``would-be 
pure gauge'' states that become physical because the symmetry is 
altered by the requirement that a horizon exist.\\

To appear in \emph{The Kerr spacetime: rotating black holes in general 
relativity}, edited by S.\ Scott, M.\ Visser, and D.\ Wiltshire (Cambridge 
University Press).
}
\end{minipage}
\end{center}
\end{titlepage}
\addtocounter{footnote}{-1}

\section{Introduction}
 
It has been more than thirty years since Bekenstein \cite{Bekenstein}
and Hawking \cite{Hawking} first taught us that black holes are 
thermodynamic objects, with characteristic temperatures
\begin{equation}
T_{\scriptstyle\mathit{H}} = \frac{\hbar\kappa}{2\pi c}
\label{ca1}
\end{equation}
and entropies
\begin{equation}
S_{\scriptstyle\mathit{BH}} = \frac{A}{4\hbar G} ,
\label{ca2}
\end{equation}
where $\kappa$ is the surface gravity and $A$ is the area of the event 
horizon.  Extensive experience with thermodynamics in less exotic settings 
encourages us to believe that these quantities should reflect some kind 
of underlying statistical mechanics.  The Bekenstein-Hawking entropy 
(\ref{ca2}), for example, should count the number of microscopic states 
of the black hole.  But by Wheeler's famous dictum, ``a black hole has no 
hair'': a classical, equilibrium black hole is determined completely by 
its mass, charge, and angular momentum, with no room for additional 
microscopic states to account for thermal behavior.  

If black hole thermodynamics has a statistical mechanical origin, the 
relevant states must therefore be nonclassical.  Indeed, they should be 
quantum gravitational---the Hawking temperature and Bekenstein-Hawking 
entropy depend upon both Planck's constant $\hbar$ and Newton's constant $G$.  
Thus the problem of black hole statistical mechanics is not just a technical 
question about some particular configurations of matter and gravitational
fields; if we are lucky, it may give us new insight into the profound 
mysteries of quantum gravity.

This chapter will focus on an attempt to find a ``universal'' description of
black hole statistical mechanics, one that involves quantum gravity but does
not depend on fine details of any particular model of quantum gravity.
This work is incomplete and tentative, and might ultimately prove to be
wrong.  But even if it is wrong, my hope is that we will learn something 
of value along the way.

\section{Black hole entropy and the problem of universality \label{univ}}

Ten years ago, the question of what microstates were responsible for 
black hole thermodynamics would have met with an almost unanimous answer: 
``We don't know.''  There were interesting ideas afloat, involving entanglement 
entropy \cite{Sorkin} (the entropy coming from correlations between states 
inside and outside the horizon) and entropy of an ``atmosphere'' of external 
fields near the horizon \cite{tHooft}, but we had nothing close to a complete 
description.

Today, by contrast, we suffer from embarrassment of riches.  We have many
candidates for the microscopic states of a black hole, all different, but
all apparently giving the same result \cite{Wald}.  In particular, black hole 
entropy may count:

\begin{itemize}
\item Weakly coupled string and D-brane states \cite{StromVafa,Peet}:
 black holes can be constructed in semiclassical string theory as bound states 
 of strings and higher-dimensional D-branes.  For supersymmetric (BPS) 
 configurations, thermodynamic properties computed at weak couplings 
 are protected by symmetries as one dials couplings up to realistic values,
 so weakly coupled states can be counted to determine the entropy.
\item Nonsingular geometries \cite{Mathur}: The weakly coupled D-brane excitations
 that account for black hole entropy in string theory may correspond to certain 
 nonsingular, horizonless geometries; a typical black hole state would then be 
 a ``fuzzball'' superposition of such states.
\item States in a dual conformal field theory ``at infinity'' \cite{AGMOO}:
 for black holes whose near-horizon geometry looks like anti-de Sitter space,
 AdS/CFT duality can be used to translate questions about thermodynamics to
 questions in a lower-dimensional dual, nongravitational conformal field theory.
\item Spin network states crossing the horizon \cite{Ashtekar}: in loop
 quantum gravity, one can isolate a boundary field theory at the horizon and 
 relate its states to the states of spin networks that ``puncture'' the 
 horizon.  The entropy apparently depends on one undetermined parameter, but
 once that parameter is fixed for one type of black hole, the approach yields
 the correct entropy for a wide range of other black holes.
\item ``Heavy'' degrees of freedom in induced gravity \cite{Fursaev}: as
 Sakharov first suggested \cite{Sakharov}, the Einstein-Hilbert action can be 
 induced in a theory with no gravitational action by integrating out heavy 
 matter fields.  The Bekenstein-Hawking entropy can then be computed by counting 
 these underlying  massive degrees of freedom.
\item No local states \cite{Hawkingb}: in the Euclidean path integral approach, 
 black hole thermodynamics is determined by global topological features of
 spacetime rather  than any local properties of the horizon.  Perhaps no localized
 degrees of freedom are required to account for black hole entropy.
\item Nongravitational states: Hawking's original calculation of black hole
 radiation was based on quantum field theory in a fixed, classical black hole
 background.  Perhaps the true degrees of freedom are not gravitational at all,
 but represent ``entanglement entropy'' \cite{Sorkin} or the states of matter near
 the horizon \cite{tHooft}.
\end{itemize}

None of these pictures has yet given us a complete model for black hole
thermodynamics.  But each can be used to count states for at least one class 
of black holes, and within its realm of applicability, each seems to give the correct 
Bekenstein-Hawking entropy (\ref{ca2}).  In an open field of investigation, the 
existence of competing explanations may be seen as a sign of health.  But the 
existence of competing explanations that all \emph{agree} is also, presumably, 
a sign that we are missing some deeper underlying structure.  

This problem of universality occurs even within particular approaches to 
black hole statistical mechanics.  In string theory, for example, one does not 
typically relate black hole entropy directly to horizon area.  Rather, one 
constructs a particular semiclassical black hole as an assemblage of strings 
and D-branes; computes the entropy at weak coupling as a function of various 
charges; and then, separately, computes the horizon area as a function of 
those same charges.  The results agree with the Bekenstein-Hawking formula 
(\ref{ca2}), but the agreement has to be checked case by case.  We may ``know'' 
that the next case will agree as well; but in a deep sense, we do not know why.

A simple approach to this problem is to note that black hole temperature and 
entropy can be computed semiclassically, so any quantum theory of gravity 
that has the right classical limit will have to give the ``right'' answer.  
But while this may be true, it does not really address the fundamental question: 
how is it that the classical theory enforces these restrictions on a quantum 
theory?  In ordinary thermodynamics, we can determine entropy classically, by 
computing the volume of the relevant region of phase space; the correspondence 
principle then ensures that the quantum mechanical answer will be the same to 
lowest order.  For black holes, no such calculation seems possible: once the mass,
charge, and spin are fixed, there is no classical phase space left.  Once again,
we are missing a vital piece of the puzzle.

\section[Symmetries and state-counting]{Symmetries and state-counting:
conformal field theory and the Cardy formula \label{SSC}}

One possibility is that the missing piece is a classical symmetry of black hole
spacetimes.  Such a symmetry would be inherited by any quantum theory, regardless
of the details of the quantization.   At first sight, this explanation seems 
unlikely: we are not used to the idea that a symmetry can be strong enough to
determine such detailed properties of a quantum theory as the density of states.  
In at least one instance, though, this is known to happen.

Consider a two-dimensional conformal field theory, defined initially on the
complex plane with coordinates $(z,{\bar z})$.  The holomorphic diffeomorphisms
$z\rightarrow f(z)$, ${\bar z}\rightarrow{\bar f}({\bar z})$ are symmetries 
of such a theory.  Denote by $L^{\hbox{\scriptsize\it cl}}_n$ and 
${\bar L}^{\hbox{\scriptsize\it cl}}_n$ the generators of the transformations
\begin{equation}
z\rightarrow z + \epsilon z^{n+1} , \qquad 
{\bar z}\rightarrow {\bar z} + \epsilon {\bar z}^{n+1} .
\label{cb1}
\end{equation}
(The superscript {\it cl} means ``classical.'')
The Poisson brackets of these generators are almost uniquely determined by the
symmetry: they form a Virasoro algebra,
\begin{align}
\left\{L^{\hbox{\scriptsize\it cl}}_m,L^{\hbox{\scriptsize\it cl}}_n\right\} 
  &= i(n-m)L^{\hbox{\scriptsize\it cl}}_{m+n} 
  + \frac{ic^{\hbox{\scriptsize\it cl}}}{12}n(n^2-1)\delta_{m+n,0} 
  \nonumber\\
\left\{{\bar L}^{\hbox{\scriptsize\it cl}}_m,{\bar L}^{\hbox{\scriptsize\it cl}}_n\right\} 
  &= i(n-m){\bar L}^{\hbox{\scriptsize\it cl}}_{m+n} 
  + \frac{ic^{\hbox{\scriptsize\it cl}}}{12}n(n^2-1)\delta_{m+n,0} \label{cb2}\\
\left\{L^{\hbox{\scriptsize\it cl}}_m,{\bar L}^{\hbox{\scriptsize\it cl}}_n\right\} 
  &= 0 \nonumber ,
\end{align}
where $c^{\hbox{\scriptsize\it cl}}$ is a constant, the central charge \cite{CFT}. 

When $c^{\hbox{\scriptsize\it cl}}=0$, equation (\ref{cb2}) is just a representation 
of the ordinary algebra of holomorphic vector fields,
\begin{equation}
\left[ z^{m+1}\frac{d\ }{dz}, z^{n+1}\frac{d\ }{dz} \right] 
  = (n-m)z^{m+n+1}\frac{d\ }{dz} .
\label{cb2a}
\end{equation}
Up to field redefinitions, the Virasoro algebra is the only central extension
of this algebra.  The central charge $c$ commonly appears quantum mechanically,
arising from operator reorderings.  But can occur classically as well.  There, 
it appears because the canonical generators are unique only up to the addition 
of constants; a central charge represents a nontrivial cocycle \cite{Arnold}, 
that is, a set of constants that cannot be removed by field redefinitions.  

In 1986, Cardy found a remarkable property of such conformal field theories 
\cite{Cardy,Cardyb}.  Let $\Delta_0$ be the smallest eigenvalue of $L_0$ in the 
spectrum, and define an ``effective central charge''
\begin{equation}
c_{\hbox{\scriptsize\it eff}} = c-24\Delta_0 .
\label{cb3}
\end{equation} 
Then for large $\Delta$, the density of states with eigenvalue $\Delta$
of $L_0$ has the asymptotic form
\begin{equation}
\rho(\Delta) \sim
\exp\left\{2\pi\sqrt{\frac{c_{\hbox{\scriptsize\it eff}}\Delta}{6}}\right\}
\rho(\Delta_0) ,
\label{cb4}
\end{equation}
independent of any other details of the theory.  The asymptotic behavior
of the density of states is thus determined by a few features of the symmetry,
the central charge $c$ and the ground state conformal weight $\Delta_0$.
In particular, theories with very different field contents can have exactly 
the same asymptotic density of states.

A careful proof of this result, using the method of steepest descents, is
given in \cite{Carlipa}.  One can derive the logarithmic corrections to 
the entropy by the same methods \cite{Carlipb}; indeed, by exploiting results 
from the theory of modular forms, one can obtain even higher order corrections 
\cite{Birm,Farey}.  But although the mathematical derivation of the Cardy 
formula is relatively straightforward, I do not know of any good, intuitive 
\emph{physical} explanation for (\ref{cb4}).  Standard derivations rely 
on a duality between high and low temperatures, which arises because of modular 
invariance: by interchanging cycles on a torus, one can trade a system on 
a circle of circumference $L$ with inverse temperature $\beta$ for a system 
on a circle of circumference $\beta$ with inverse temperature $L$.  Such a
transformation relates states at high ``energies'' $\Delta$ to the ground 
state, leading ultimately to Cardy's result.  But it would be valuable to have 
a more direct understanding of why the density of states should be so strongly
constrained by symmetry.

Note that upon quantization, after making the usual substitutions $\{\,,\,\}%
\rightarrow[\,,\,]/i\hbar$ and $L^{\hbox{\scriptsize\it cl}}_m \rightarrow L_m/\hbar$, 
the Virasoro algebra (\ref{cb2}) becomes
\begin{align}
\left[L_m,L_n\right] 
  &= (m-n)L_{m+n} 
  + \frac{c^{\hbox{\scriptsize\it cl}}}{12\hbar}m(m^2-1)\delta_{m+n,0} 
  \nonumber\\
\left[{\bar L}_m,{\bar L}_n\right] 
  &= (m-n){\bar L}_{m+n} 
  + \frac{c^{\hbox{\scriptsize\it cl}}}{12\hbar}m(m^2-1)\delta_{m+n,0} \label{cb2c}\\
\left[L_m,{\bar L}_n\right] 
  &= 0 \nonumber .
\end{align}
A nonvanishing classical central charge $c^{\hbox{\scriptsize\it cl}}$ thus 
contributes  $c^{\hbox{\scriptsize\it cl}}/\hbar$ to the quantum central charge, and 
a classical conformal ``charge'' $\Delta^{\hbox{\scriptsize\it cl}}$
gives a quantum conformal weight $\Delta^{\hbox{\scriptsize\it cl}}/\hbar$.
Hence the classical contribution to the entropy $\log\rho(\Delta)$ coming from 
(\ref{cb4}) goes as $1/\hbar$, matching the behavior of the Bekenstein-Hawking 
entropy (\ref{ca2}) and giving us a first hint that an approach of this sort 
might be productive.

The black holes we are interested in are not two dimensional, of course, and
despite some interesting speculation \cite{Verlinde}, there is no proven 
higher-dimensional analog to the Cardy formula.  But there is reason to hope
that the two-dimensional result (\ref{cb4}) might be relevant to the near-%
horizon region of an arbitrary black hole.  For instance, it is known that 
near a horizon, matter can be described by a two-dimensional conformal field 
theory \cite{Birmingham,Gupta}.  Indeed, in ``tortoise coordinates,'' the 
near-horizon metric in any dimension becomes
\begin{equation}
ds^2 = N^2(dt^2-dr_*{}^2) + ds_\perp{}^2 ,
\label{cb5}
\end{equation}
where the lapse function $N$ goes to zero at the horizon.  The Klein-Gordon
equation then reduces to
\begin{equation}
(\Box - m^2)\varphi =
     \frac{1}{N^2}(\partial_t^2 - \partial_{r_*}^2)\varphi + \mathcal{O}(1) = 0 .
\label{cb6}
\end{equation}
The mass and transverse excitations become negligible near the horizon: they
are essentially red-shifted away relative to excitations in the $r_*$-$t$
plane, leaving an effective two-dimensional conformal field theory at each 
point of the horizon.  A similar reduction occurs for spinor and vector fields. 
Moreover, Jacobson and Kang have observed that the surface gravity and temperature 
of a stationary black hole are conformally invariant \cite{Jacobson}, and Medved 
et al.\ have recently shown that a generic stationary black hole metric has 
an approximate conformal symmetry near the horizon \cite{Martin,Martinb}.

\section{The BTZ black hole \label{BTZ}}

The first concrete evidence that conformal symmetry can determine black hole 
thermodynamics came from studying the (2+1)-dimensional black hole of Ba{\~n}ados,
Teitelboim, and Zanelli \cite{BTZ,BHTZ,Carlipc}.  A solution of the vacuum 
Einstein equations with a negative cosmological constant $\Lambda = -1/\ell^2$,
the BTZ black hole is the (2+1)-dimensional analog of the Kerr-AdS metric.   
In Boyer-Lindquist-like coordinate, the BTZ metric takes the form
\begin{multline}
ds^2 = N^2dt^2 - N^{-2}dr^2 - r^2\left( d\phi + N^\phi dt\right)^2
\label{cc1}\\
\hbox{with}\ N 
  = \left( -8GM + \frac{r^2}{\ell^2} + \frac{16G^2J^2}{r^2} \right)^{1/2},\ \
N^\phi = - \frac{4GJ}{r^2} ,
\end{multline}
where $M$ and $J$ are the anti-de Sitter analogs of the ADM mass and angular 
momentum.  As in 3+1 dimensions, the apparent singularities at $N=0$ are
merely coordinate singularities, and an analog of Kruskal-Szekeres coordinates
can be found.

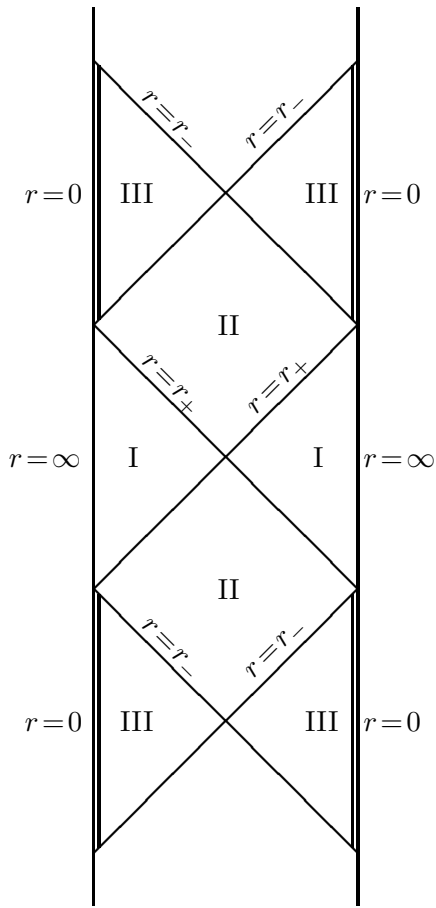
\begin{figure}
\centering
\begin{picture}(240,340)(-75,0)
\thicklines
%
%
\put(0,0){\line(0,1){340}}     
\put(100,0){\line(0,1){340}}   
\put(2,22){\line(0,1){96}}
\put(98,22){\line(0,1){96}}
\put(2,222){\line(0,1){96}}
\put(98,222){\line(0,1){96}}
\put(0,20){\line(1,1){100}}
\put(0,120){\line(1,-1){100}}
\put(0,120){\line(1,1){100}}
\put(0,220){\line(1,-1){100}}
\put(0,220){\line(1,1){100}}
\put(0,320){\line(1,-1){100}}
\put(10,66){III}
\put(80,66){III}
\put(47,116){II}
\put(13,166){I}
\put(83,166){I}
\put(47,216){II}
\put(10,266){III}
\put(80,266){III}
\put(-26,66){$r\!=\!0$}
\put(102,66){$r\!=\!0$}
\put(-32,166){$r\!=\!\infty$}
\put(102,166){$r\!=\!\infty$}
\put(-26,266){$r\!=\!0$}
\put(102,266){$r\!=\!0$}
\put(57,86){\rotatebox{45}{$r\!=\!r_-$}}
\put(58,186){\rotatebox{45}{$r\!=\!r_+$}}
\put(57,286){\rotatebox{45}{$r\!=\!r_-$}}
\put(16,106){\rotatebox{-45}{$r\!=\!r_-$}}
\put(16,206){\rotatebox{-45}{$r\!=\!r_+$}}
\put(16,306){\rotatebox{-45}{$r\!=\!r_-$}}
\end{picture}
\caption{The Carter-Penrose diagram for a nonextremal BTZ black hole
 \label{fig1}}
\end{figure}

Like all vacuum spacetimes in 2+1 dimensions, the BTZ geometry has the peculiar 
feature of having constant curvature, and can be in fact expressed as a quotient 
of anti-de Sitter space by a discrete group of isometries.  Nevertheless, it is 
a genuine black hole:
\begin{itemize}
\item It has a true event horizon at $r=r_+$ and, if $J\ne0$, an inner Cauchy
horizon at $r=r_-$, where
\begin{equation}
r_\pm^2=4GM\ell^2\left \{ 1 \pm
\left [ 1 - \left(\frac{J}{M\ell}\right )^2\right ]^{1/2}\right \} ;
\label{cc2}
\end{equation}
\item it occurs as the end point of gravitational collapse of matter;
\item its Carter-Penrose diagram, figure \ref{fig1}, is essentially the same as
 that of an ordinary Kerr-AdS black hole;  
\item and, most important for our purposes, it exhibits standard black hole 
thermodynamics, with a temperature and entropy given by (\ref{ca1}) and (\ref{ca2}), 
where the horizon ``area'' is the circumference $A=2\pi r_+$.
\end{itemize}

The thermodynamic character of the BTZ black hole may be verified in much 
the same way that it is for the Kerr black hole: by looking at quantum field 
theory in a BTZ background \cite{Lif,Hyun}; by examining the Euclidean path integral 
\cite{CarTeit} and the Brown-York microcanonical path integral \cite{BMann}; 
by appealing to Wald's Noether charge approach \cite{Carlipc,CGeg}; and by
investigating tunneling through the horizon \cite{EngRez,Medved}.  In 2+1 
dimensions, a powerful new method is also available \cite{Emparan}: one can 
consider quantum gravitational perturbations induced by a classical scalar 
source, and then use detailed balance arguments to obtain thermodynamics.
The quantitative agreement of all of these approaches gives us confidence that 
the thermal properties are real.

We now come to a deep mystery.  Vacuum general relativity in 2+1 dimensions
has no dynamical degrees of freedom.  This is most easily seen by a simple 
counting argument---in the canonical formalism, the field is described by
a spatial metric (three degrees of freedom per point) and its canonical
momentum (three more degrees of freedom per point), but we also have three
constraints that restrict initial values and three arbitrary coordinate 
choices, leaving $6-6=0$ dynamical degrees of freedom.  The same conclusion
may be reached by noting that the (2+1)-dimensional curvature tensor is
algebraically determined by the Einstein tensor,
\begin{equation}
G^\mu{}_\nu = -\frac{1}{4}\epsilon^{\mu\pi\rho}\epsilon_{\nu\sigma\tau}
  R_{\pi\rho}{}^{\sigma\tau} ,
\label{cc2a}
\end{equation}
so a spacetime that is empty apart from the presence of a cosmological 
constant necessarily has constant curvature.  While the theory admits a 
few global ``topological'' excitations \cite{Carlipd}, there is no local 
dynamics.  Where, then, can the degrees of freedom responsible for 
thermal behavior come from?

One piece of the answer, discovered independently by Strominger \cite{Strom}
and Birmingham, Sachs, and Sen \cite{BSS}, can be found by looking at boundary
conditions at infinity.  The conformal boundary of a (2+1)-dimensional
asymptotically anti-de Sitter spacetime is a cylinder, so it is not surprising
that the asymptotic symmetries of the BTZ black hole are described by a Virasoro 
algebra (\ref{cb2}).  It is a bit more surprising that this algebra has a 
central extension, but as Brown and Henneaux showed in 1986 \cite{BH}, the 
classical central charge, computed from the standard ADM constraint algebra, 
is nonzero:
\begin{equation}
c = \frac{3\ell}{2G} .
\label{cc3}
\end{equation}
Confirmation of this result has come from a path integral analysis 
\cite{Terashima}, from investigating the constraint algebra in the Chern-Simons 
formalism \cite{Banados}, and from examining the conformal anomaly of the 
boundary stress-energy tensor \cite{HennSken,BalKraus}.  

Moreover, the classical Virasoro ``charges'' $L_0$ and ${\bar L}_0$ can be 
computed within ordinary canonical general relativity, employing the same 
techniques that are used to determine the ADM mass \cite{BH}.  Indeed, in this 
context the zero-modes of the diffeomorphisms (\ref{cb1}) are simply linear 
combinations of time translations and rotations, and the corresponding conserved 
quantities are linear combinations of the ordinary ADM mass and angular momentum.  
For the BTZ black hole, in particular, one finds
\begin{equation}
\Delta = \frac{1}{16G\ell} (r_+ + r_-)^2, \quad 
{\bar\Delta} = \frac{1}{16G\ell} (r_+ - r_-)^2 .
\label{cc4}
\end{equation}
By the Cardy formula (\ref{cb4}), with the added assumption that $\Delta_0$
and ${\bar\Delta}_0$ are small, one then obtains an entropy
\begin{equation}
S = \log\rho\sim \frac{2\pi}{8G}(r_+ + r_-) + \frac{2\pi}{8G}(r_+ - r_-) 
  = \frac{2\pi r_+}{4G} ,
\label{cc5}
\end{equation}
in precise agreement with the Bekenstein-Hawking entropy.

This argument is incomplete, of course: it tells us that the entropy is
related to symmetries and boundary conditions at infinity, but does not 
explain the underlying quantum degrees of freedom.  I will argue later that 
this is a \emph{good} feature, since it allows us to explain the ``universality''
described in section \ref{univ}.  For the BTZ black hole, though, we can
go a bit further.  

As first observed by Achucarro and Townsend \cite{Achucarro} and subsequently 
extensively developed by Witten \cite{Witten,Wittena}, vacuum Einstein gravity 
in 2+1 dimensions with a negative cosmological constant is equivalent 
to a Chern-Simons gauge theory, with a gauge group $\mathrm{SL}(2,\mathbb{R})%
\times\mathrm{SL}(2,\mathbb{R})$.  On a compact manifold, a Chern-Simons
theory is a ``topological field theory,'' described by a finite number of
global degrees of freedom.  On a manifold with boundary, however, boundary
conditions can partially break the gauge invariance.  As a consequence, field 
configurations that would ordinarily be considered gauge-equivalent become
distinct at the boundary.  In a manner reminiscent of the Goldstone mechanism 
\cite{Kaloper}, new ``would-be pure gauge'' degrees of freedom appear, providing
new dynamical degrees of freedom.   

For a Chern-Simons theory, the resulting induced boundary dynamics can be
described by a Wess-Zumino-Witten model \cite{Wittenb,EMSS}.  For (2+1)-dimensional
gravity, slightly stronger boundary conditions can further reduce the boundary theory
to Liouville theory \cite{CHvD,Rooman}, a result that may also be obtained directly
in the metric formalism \cite{SkenSolo,Bautier,Roomanb,Carlipe}.  Whether the 
resulting degrees of freedom reproduce the Bekenstein-Hawking entropy remains an 
open question---for a recent review, see \cite{Carlipf}---but Chen has found 
strong hints that a better understanding of Liouville theory might allow an 
explicit microscopic description of BTZ black hole thermodynamics \cite{Chen}.

\section{Horizon constraints}

The BTZ black hole offers a test case for the hypothesis that black hole entropy
might be controlled by an underlying classical symmetry.  But it is clearly not
good enough.  To start with, the computations described in the preceding section 
relied heavily on a very particular feature of (2+1)-dimensional asymptotically 
anti-de Sitter spacetime, the fact that the ``boundary'' at which asymptotic 
diffeomorphisms are defined is a two-dimensional timelike surface.  A few other 
black holes have a similar character: the near-extremal black holes considered 
in string theory often have a near-horizon geometry that looks like that of a 
BTZ black hole, and two-dimensional methods can be used to obtain their entropy 
\cite{Strom,Skenderis}.  But for the generic case, there is no reason to expect 
such a nice structure.

Moreover, the standard computations of BTZ black hole entropy use conformal 
symmetries at infinity.  For a single, isolated black hole in 2+1 dimensions, 
this choice is probably harmless: there are no propagating degrees of freedom 
between the horizon and infinity, so it may not matter where we count the states.  
But even in 2+1 dimensions, the interpretation becomes unclear when there is more 
than one black hole present, or when the black hole is replaced by a ``star'' for 
which the BTZ solution is only the exterior geometry.  If we wish to isolate the 
microscopic states of a particular black hole, it will be very difficult to do
so using only symmetries at conformal infinity; we should presumably be looking 
near the horizon instead.

We are thus left with several hints.  We should
\begin{itemize}
\item Look for ``broken gauge invariance'' to provide new degrees of freedom;
\item Hope for an effective two-dimensional picture, which would allow us to
use the Cardy formula;
\item But look near horizon.
\end{itemize}
Before proceeding, though, we need to take a step back and ask a more general
question.  We want to investigate the statistical mechanics of a black hole.
But how, exactly, do we tell that a black hole is present in the context of
a fully quantum mechanical theory of gravity?

This question is frequently overlooked, because in the usual semiclassical
approaches to black hole thermodynamics the answer is obvious: we fix a definite
black hole background and then ask about quantum fields and gravitational
fluctuations in that background.  But in a full quantum theory of gravity, we 
cannot do that: there is no fixed background, the geometry is quantized, and
the uncertainty principle prevents us from exactly specifying the metric.
In general, it becomes difficult to tell whether a black hole is present or
not.  At best, we can ask a \emph{conditional} question: ``If a black hole with
characteristic X is present, what is the probability of phenomenon Y?''

There are two obvious ways to impose a suitable condition to 
give such a conditional probability.  One, discussed in \cite{Carlipg}, 
is to treat the horizon as a ``boundary'' and impose appropriate black hole 
boundary conditions.  The horizon is not, of course, a true boundary: a 
falling observer can cross a horizon without seeing anything special 
happen, and certainly doesn't drop off the edge of the Universe.  Nonetheless, 
we can ensure the presence of a black hole by specifying ``boundary conditions'' 
at a horizon.  In a path integral formulation, for example, we can divide the 
manifold into two pieces along a hypersurface $\Sigma$ and perform separate 
path integrals over the fields on each piece, with fields restricted at the 
``boundary'' by the requirement that $\Sigma$ be a suitable black hole horizon.  
This kind of split path integral has been studied in detail in 2+1 dimensions 
\cite{Wittenc}, where it yields the correct counting for the boundary degrees 
of freedom.

Alternatively, we can impose ``horizon constraints'' directly, either classically
or in the quantum theory.  We might, for example, construct an operator $\vartheta$
representing the expansion of a particular null surface, and restrict ourselves 
to states annihilated by $\vartheta$.  As we shall see below, such a restriction 
can alter the algebra of diffeomorphisms, allowing us to exploit the Cardy formula.

The ``horizon as boundary'' approach has been explored by a number of authors; see,
for instance, \cite{Carliph,Carlipi,Navarro,Jing,Izq,Park,Silva,Cvitan,Cvitanb}.  A 
conformal symmetry in the $r$--$t$ plane appears naturally, and one can obtain a 
Virasoro algebra with a central charge that leads to the correct Bekenstein-Hawking 
entropy.  But the diffeomorphisms whose algebra yields that central charge---%
essentially, the diffeomorphisms that leave the lapse function invariant---are 
generated by vector fields that blow up at the horizon \cite{Dreyer,Koga,Pinamonti}, 
and it is not clear whether they should be permitted in the theory.  A closely
related approach looks for an approximate conformal symmetry in the neighborhood 
of the horizon \cite{Solo,Carlipj,Giacomini,Solob}.  Again, one finds a Virasoro 
algebra with a central charge that apparently leads to the correct Bekenstein-Hawking 
entropy.

The ``horizon constraint'' approach is much newer, and is not yet fully developed.  
Let us now examine it further.  Suppose we wish to constrain our theory of gravity 
by requiring that some surface $\Sigma$ be the horizon of a black hole.  We must 
first decide exactly what we mean by a ``horizon.''  Demanding that $\Sigma$ be a 
true event horizon seems impractical: an event horizon is determined globally, 
and requires that we know the entire future development of the spacetime.  The 
most promising alternative is probably offered by the ``isolated horizon'' 
program \cite{Ashtekarb}.  An isolated horizon is essentially a null surface with 
vanishing expansion, with a few added technical conditions.  Such a horizon shares 
many of the fundamental features of an event horizon \cite{Ashtekarc}---in particular, 
it leads to standard black hole thermodynamics---and seems to do a good job of
capturing the idea of a ``local'' horizon.

Isolated horizon constraints are constraints on the allowed data on a hypersurface,
and in principle we should be able to use the well-developed apparatus of constrained
Hamiltonian dynamics \cite{Dirac,Diracb,Bergmann} to study such conditions.  
Unfortunately, though an isolated horizon is by definition a null surface, and
we would require an approach akin to light cone quantization.  Light cone quantization 
of gravity is difficult, and at this writing, such an analysis of horizon constraints 
has not been carried out.  

\begin{figure}
\centering
\begin{picture}(200,180)(32,-10)
\thicklines
%
%
\put(50,0){\line(1,1){160}}
\put(50,160){\line(1,-1){160}}
\put(50,0){\line(-1,1){80}}
\put(-30,80){\line(1,1){80}}
\put(210,0){\line(1,1){80}}
\put(290,80){\line(-1,1){80}}
\multiput(50,160)(20,0){8}{\line(1,1){10}}
\multiput(60,170)(20,0){8}{\line(1,-1){10}}
\multiput(50,0)(20,0){8}{\line(1,-1){10}}
\multiput(60,-10)(20,0){8}{\line(1,1){10}}
\qbezier(130,80)(145,85)(214,154)
\put(173,105){$\Delta$}
\end{picture}
\caption{A spacelike ``stretched horizon'' $\Delta$ \label{fig2}}
\end{figure}
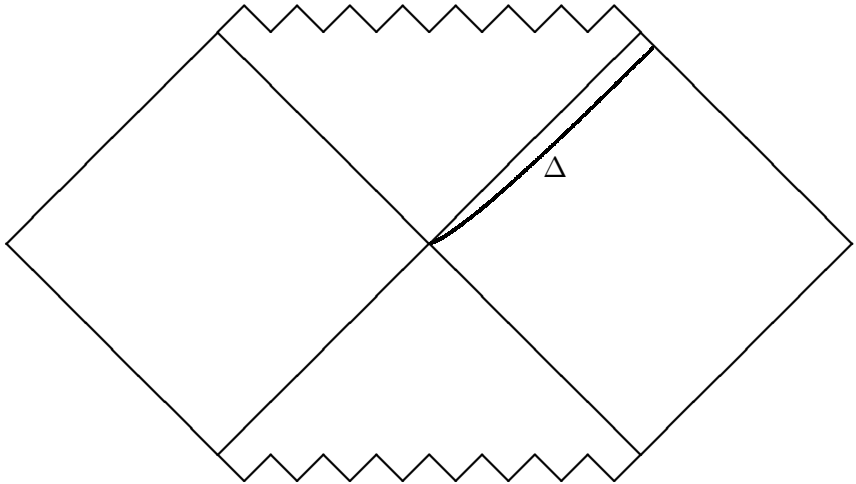

We can much more easily impose constraints requiring the presence of a spacelike 
``stretched horizon'' that becomes nearly null, as illustrated in figure \ref{fig2}.  
On such a hypersurface, it is possible to employ standard methods of constrained 
dynamics.  As we shall see below, at least for the relatively simple model of 
two-dimensional dilaton gravity, the horizon constraints lead to a Virasoro algebra 
with a calculable central charge, allowing us to use the Cardy formula to obtain 
the correct Bekenstein-Hawking entropy.

\section{Dilaton black holes and the Bekenstein-Hawking entropy \label{DBH}}

As noted in section \ref{SSC}, it is plausible that the thermodynamic properties 
of a black hole are determined by dynamics in the ``$r$--$t$ plane.''  As 
a first step, we can therefore look at a dimensionally reduced theory.  The full
Einstein-Hilbert action in any dimension may be written exactly as an action for
a two-dimensional theory: standard Kaluza-Klein techniques allow us to express 
the effect of ``extra dimensions'' in terms of scalar and gauge fields, albeit 
with an enormous gauge group \cite{Yoon}.  Near a black hole horizon, though, 
most of these fields become negligible, and one obtains a simple two-dimensional 
theory, dilaton gravity \cite{Kunstatter,Grumiller}, with an action of the form
\begin{equation}
I = \frac{1}{16\pi G}\int d^2x \sqrt{-g}\left[AR + V(A)\right] .
\label{ce1}
\end{equation}
Here, $R$ is the two-dimensional scalar curvature, while $A$ is a scalar field, 
the dilaton (often denoted $\varphi$).  $V(A)$ is a potential whose detailed form 
depends on the higher-dimensional theory we started with; we will not need an 
exact expression. As the notation suggests, $A$ is the transverse area in the 
higher-dimensional theory, and the expansion---the fractional rate of change of 
the transverse area along a null curve with null normal $l^a$---becomes
\begin{equation}
\vartheta = l^a\nabla_aA/A .
\label{ce1a}
\end{equation}

It is useful to reexpress the action (\ref{ce1}) in terms of a null dyad $(l^a,n^a)$
with $l^2=n^2=0$, $l\cdot n=-1$.  These determine ``surface gravities'' $\kappa$ and 
$\bar\kappa$, defined by the conditions
\begin{align}
&\nabla_al_b = -\kappa n_al_b - {\bar\kappa}l_al_b \nonumber\\
&\nabla_an_b = \kappa n_an_b + {\bar\kappa}l_an_b .
\label{ce2}
\end{align}
By an easy computation, the action (\ref{ce1}) becomes  
\begin{equation}
I =  \int d^2x \left[{\hat\epsilon}^{ab}
     \left(2\kappa n_b\partial_aA
     - 2{\bar\kappa} l_b\partial_aA\right) + \sqrt{-g}V\right] ,
\label{ce3}
\end{equation}
where ${\hat\epsilon}^{ab}$ is the two-dimension Levi-Civita density and I have
adopted units such that $16\pi G=1$.  If we now define components of our dyad
with respect to coordinates $(u,v)$,
\begin{equation}
l = \sigma du + \alpha dv, \qquad n = \beta du + \tau dv ,
\label{ce4}
\end{equation}
it is straightforward to find the Hamiltonian form of the action.  Details may be 
found in \cite{Carlipk}.  The key feature is that the system contains three 
first-class constraints,
\begin{align} 
&C_\perp = \pi_\alpha{}' - \frac{1}{2}\pi_\alpha\pi_A - \tau V(A) 
 \nonumber\\
&C_\parallel = \pi_A A' - \alpha\pi_\alpha{}' - \tau\pi_\tau{}' \vphantom{\frac{1}{2}}
 \label{ce5}\\
&C_\pi = \tau\pi_\tau - \alpha\pi_\alpha + 2A' \vphantom{\frac{1}{2}},\nonumber
\end{align}
where a dot denotes a derivative with respect to $u$, a prime a derivative
with respect to $v$, and $\pi_X$ is the momentum conjugate to the field $X$.  
$C_\perp$ and $C_\parallel$ are ordinary Hamiltonian and momentum constraints
of general relativity, that is, the canonical versions of the generators of 
diffeomorphisms; $C_\pi$ is a disguised version of the generator of local Lorentz 
invariance, appearing because the pair $\{l,n\}$ is invariant under the boost 
$l\rightarrow fl$, $n\rightarrow f^{-1}n$.  

We can now impose our ``stretched horizon'' constraints at the surface $\Sigma$
defined by the condition $u=0$.  We first demand that $\Sigma$ be ``almost 
null,'' i.e., that its normal be nearly equal to the null vector $l^a$.  By
(\ref{ce4}), this requires that $\alpha=\epsilon_1\ll1$.  

We next demand that $\Sigma$ be ``almost nonexpanding,'' that is, that the
expansion $\vartheta$ be ``almost zero'' on $\Sigma$.  This condition is slightly
more subtle, since the absolute scale of $l^a$ is not fixed.  While the
restriction $\vartheta=0$ is independent of this scale, a restriction of the
form $\vartheta\ll1$ clearly is not.  Fortunately, though, the surface gravity
$\kappa$ scales identically under constant rescalings of $l^a$, so we can
consistently require that $l^v\nabla_vA/\kappa A = \epsilon_2\ll1$.  Rewriting
these conditions in terms of canonical variables, we obtain two constraints:
\begin{align}
&K_1 = \alpha-\epsilon_1 = 0 \nonumber\\
&K_2 = A' - \frac{1}{2}\epsilon_2A_+\pi_A + \frac{a}{2}C_\pi = 0 ,
\label{ce6}
\end{align}
where $A_+$ is the value of the dilaton at the horizon.  The term proportional 
to $C_\pi$ in $K_2$ is not necessary, but has been added for later convenience.
By looking at a generic exact solution, one can verify that for $\epsilon_2<0$,  
these constraints do, in fact, define a spacelike stretched horizon of the type
illustrated in figure \ref{fig2}.

$K_1$ and $K_2$ are not quite ``constraints'' in the ordinary sense of constrained
Hamiltonian dynamics.  In particular, they restrict allowable data only on the 
stretched horizon $\Sigma$, and cannot be imposed elsewhere.  Nevertheless, they 
are similar enough to conventional constraints that many existing techniques can 
be used.  In particular, note that the $K_i$ have nontrivial brackets with the 
momentum and boost generators $C_\parallel$ and $C_\pi$, so these can no longer 
be considered generators of invariances of the constrained theory.\footnote{The 
$K_i$ have nontrivial brackets with $C_\perp$ as well, but that is not a problem: 
$C_\perp$ generates diffeomorphisms that move us off the initial surface $\Sigma$, 
and we should not expect the horizon constraints to be preserved.}  But we can 
fix this in a manner suggested by Bergmann and Komar many years ago \cite{Bergmann}: 
we define new generators
\begin{align}
C_\parallel &\rightarrow C_\parallel^* = C_\parallel + a_1K_1 + a_2K_2 \nonumber\\
C_\pi &\rightarrow C_\pi^* = C_\pi + b_1K_1 + b_2K_2 
\label{ce7}
\end{align}
with coefficients $a_i$ and $b_i$ chosen so that $\{C^*,K_i\}=0$.  Since the $K_i$
vanish on admissible geometries---those for which our initial surface is a suitable
stretched horizon---the constraints $C^*$ are physically equivalent to the original
$C$; but they now preserve the horizon constraints as well.   

We now make the crucial observation that the redefinitions (\ref{ce7}) affect the 
Poisson brackets of the constraints.  With the choice $a=-2$ in (\ref{ce6}), it
is not hard to check that
\begin{align}
\{C_\parallel^*[\xi],C_\parallel^*[\eta]\} &= -C_\parallel^*[\xi\eta'-\eta\xi'] 
  + \frac{1}{2}\epsilon_2A_+\int\,dv(\xi'\eta'' - \eta'\xi'') \nonumber\\
\{C_\parallel^*[\xi],C_\pi^*[\eta]\} &= -C_\pi^*[\xi\eta'] \label{ce8}\\
\{C_\pi^*[\xi],C_\pi^*[\eta]\} &= -\frac{1}{2}\epsilon_2A_+\int\,dv(\xi\eta' - \eta\xi') 
  \nonumber
\end{align}
where $C[\xi]$ means $\int\!dv\,\xi C$.  The algebra (\ref{ce8}) has a simple conformal 
field theoretical interpretation \cite{CFT}: the $C_\parallel^*$ generate a Virasoro 
algebra with central charge
\begin{equation}
\frac{c}{48\pi} = -\frac{1}{2}\epsilon_2A_+ ,
\label{ce9}
\end{equation}
while $C_\pi^*$ is an ordinary primary field of weight one.

To take advantage of the Cardy formula (\ref{cb4}), the central charge (\ref{ce9}) 
is not enough; we also need the classical Virasoro ``charge'' $\Delta$.  As in 
conventional approaches to black hole mechanics, this charge comes from the 
contribution of a boundary term that must be added to make the Virasoro generator 
$C_\parallel^*$ ``differentiable'' \cite{Regge}.  Under a variation of the fields, 
the momentum constraint (\ref{ce5}) picks up a boundary term from integration by 
parts,
\begin{equation}
\delta C_\parallel[\xi] = \dots + \left.\xi\pi_A\delta A\right|_{v=v_+} ,
\label{ce9a}
\end{equation}
at the boundary $v=v_+$.  For Poisson brackets with $C_\parallel$ to be well-defined,
this term must be canceled.  We therefore add a boundary term to $C_\parallel$,
\begin{equation}
C_{\parallel\,\mathit{bdry}}^*[\xi] = -\left.\xi\pi_AA\right|_{v=v_+} ,
\label{ce10}
\end{equation}
which will give a nonvanishing classical contribution to $\Delta$.

We also need a ``mode expansion'' to define the Fourier component $L_0$, or,
equivalently, a normalization for the ``constant translation'' $\xi_0$.  For a
conformal field theory defined on a circle, or on a full complex plane with a
natural complex coordinate, the mode expansion (\ref{cb1}) is essentially unique.
Here, though, it is not so obvious how to choose a coordinate $z$.  As argued in
\cite{Carlipj}, however, there is one particularly natural choice,
\begin{equation}
z = e^{2\pi i A/A_+}, \qquad \xi_n = \frac{A_+}{2\pi A'} z^n,
\label{ce11}
\end{equation}
where the normalization is chosen so that $[\xi_m,\xi_n] = i(n-m)\xi_{m+n}$.

Equation (\ref{ce10}) then implies that
\begin{equation}
\Delta = C_{\parallel\,\mathit{bdry}}^*[\xi_0] = -\frac{A_+}{2\pi A'}\pi_AA_+
 = -\frac{A_+}{\pi\epsilon_2} ,
\label{ce12}
\end{equation}
where I have used the constraint $K_2=0$ to eliminate $\pi_A$ in the last equality.
Inserting (\ref{ce9}) and (\ref{ce12}) into the Cardy formula, assuming that
$\Delta_0$ is small, and restoring the factors of $16\pi G$ and $\hbar$, we 
obtain an entropy
\begin{equation}
S = \frac{2\pi}{16\pi G}\sqrt{\left(-\frac{24\pi\epsilon_2A_+}{6\hbar}\right)
  \left(-\frac{A_+}{\pi\epsilon_2\hbar}\right)} = \frac{A_+}{4\hbar G} ,
\label{ce13}
\end{equation}
exactly reproducing the Bekenstein-Hawking entropy (\ref{ca2}).

\section{What are the states?}

I argued in section \ref{univ} that one of the main strengths of a symmetry-based 
derivation is that it is ``universal,'' that is, that it does not depend on the 
details of a quantum theory of gravity.  Nevertheless, such a derivation does
allow us to say \emph{something} about the relevant states.  

Standard approach to canonical gravity require that physical states satisfy the
condition
\begin{equation}
C_\parallel|\mathit{phys}\rangle = C_\pi|\mathit{phys}\rangle = 0 ,
\label{cf1}
\end{equation}
that is, that they be annihilated by the constraints.  But the condition 
(\ref{cf1}) is not consistent with a Virasoro algebra with nonvanishing central 
charge: schematically,
\begin{equation}
[C^*,C^*]|\mathit{phys}\rangle \sim C^*|\mathit{phys}\rangle
  + \mathit{const.}|\mathit{phys}\rangle \ne 0.
\label{cf2}
\end{equation}
We must therefore weaken the physical state condition, for example by requiring
only that
\begin{equation}
C^*_\parallel{}^{(+)}|\mathit{phys}\rangle = 0  
\label{cf3}
\end{equation}
where $C^*_\parallel{}^{(+)}$ is the positive-frequency component of the momentum
constraint.  

It is well known in conformal field theory that such a loosening of the constraints 
leads to a collection of new ``descendant'' states $L_{-n}|\mathit{phys}\rangle$,
which would be excluded by the stronger constraint (\ref{cf1}).  This phenomenon
is closely analogous to the appearance of ``boundary states'' for the BTZ black 
hole, as discussed in section \ref{BTZ}.  By relaxing the physical state constraints, 
we have allowed states that would formerly have been considered to be gauge-equivalent
to differ physically, thus introducing a new set of ``would-be pure gauge'' states 
into our state-counting.  It is an interesting open question whether the 
``Goldstone-like'' description of section \ref{BTZ} can be extended to this setting.

\section{Where do we go from here?}

The ``horizon constraint'' program described in section \ref{DBH} seems promising.
But it is clear that important pieces are still missing.  In particular:
\begin{itemize}
\item Neither the constraints (\ref{ce6}) nor the mode choice (\ref{ce11}) are unique.
 It is important to understand how sensitive the final results are to these choices.
 Some flexibility certainly exists---for example, it can be shown that the final 
 expression for the black hole entropy does not depend on the parameter $a$ in 
 (\ref{ce6})---but a good deal of unexplored freedom remains.
\item At this writing, the analysis has been completed only for the two-dimensional
 dilaton black hole.  While it may be argued that this case captures the essential
 features of an arbitrary black hole, and explicit check of this claim is clearly
 needed.
\item While the final expression (\ref{ce13}) for the entropy is well-behaved, and
 exists in the limit that the ``stretched horizon'' approaches the true horizon,
 several intermediate quantities---including the central charge $c$, the conformal
 weight $\Delta$, and the vector fields $\xi_n$---behave badly at the horizon.
 Of course, the canonical approach used in section \ref{DBH} itself only makes sense
 on a spacelike stretched horizon, so this breakdown is not necessarily a sign of
 an underlying problem.  But it would clearly be desirable to perform a similar
 analysis in light cone quantization on a true horizon.
\end{itemize}
This last problem is reminiscent of the ``horizon as boundary'' approach of, for
instance, \cite{Carlipi}, in which the relevant diffeomorphisms are generated by 
vector fields that blow up at the horizon.  If this proves to be a general feature 
of conformal symmetry methods, it could be telling us something about ``black hole
complementarity'' \cite{Susskind}: perhaps the Bekenstein-Hawking entropy is only
well-defined for an observer who remains outside the horizon.

Beyond these particular issues, several more general questions must be
answered before the conformal symmetry program can be taken too seriously.  First,
we need to understand much more about the coupling of horizon degrees of freedom
to matter.  Black hole thermodynamics is, after all, more than the Bekenstein-%
Hawking entropy; one must also demonstrate that any putative horizon degrees of 
freedom couple in a way that explains Hawking radiation.  In this matter, the
string theory computations clearly have the lead, reproducing not only the Hawking
temperature but the correct gray body factors as well \cite{Peet,MalStrom}.  But 
there are a few hints that similar results can be obtained from more general
conformal symmetries \cite{Emparan,Larsen}.

Second, if the contention that near-horizon symmetry is ``universal'' is correct,
then such a symmetry must be present---albeit, perhaps, hidden---in other
derivations of black hole thermodynamics.  There are a few cases in which
this is known to be true: for example, certain string theoretical derivations
of near-extremal black hole thermodynamics exploit the conformal symmetry of
the near-horizon BTZ geometry \cite{Strom,Skenderis}, and the induced gravity
approach can be related to a two-dimensional conformal symmetry \cite{Frolovb}.
But huge gaps still remain.  One interesting avenue would be to explore the 
Euclidean version of the horizon constraint program, perhaps allowing us to 
relate the symmetry-derived count of states more directly to the Euclidean path 
integral.  For the Euclidean black hole, the ``stretched horizon'' constraints 
(\ref{ce6}) have a nice geometric interpretation, defining a circle in imaginary 
time around the horizon with a proper radius proportional to $\epsilon_2$.  It
may be that some of the choices we have made in defining the constraints become
more natural in such a setting.

Third, we will eventually have to move away from ``isolated horizons'' to
consider dynamical black holes, with horizons that grow as matter falls in
and shrink as Hawking radiation carries away energy.  It is only in such a
setting that the horizon constraint program will be able to analyze such
crucial problems as the information loss paradox and the final fate of an
evaporating black hole.  Whether this will eventually be possible remains to
be seen.

\begin{flushleft}
\bf Acknowledgments
\end{flushleft}
This work was supported in part by the U.S.\ Department of Energy under grant
DE-FG02-99ER40674.

\end{document}